\documentclass[aps,prl,twocolumn,superscriptaddress,showpacs,preprintnumbers,amsmath,amssymb]{revtex4}
\usepackage{graphicx} 
\usepackage{dcolumn}  
\usepackage{color}
\graphicspath{{ps}}

\begin{document}



\title{ \quad\\[1.0cm] Observation of the decay $B^0 \to J/\psi \ \eta$}

\affiliation{Budker Institute of Nuclear Physics, Novosibirsk}
\affiliation{Chiba University, Chiba}
\affiliation{Chonnam National University, Kwangju}
\affiliation{University of Cincinnati, Cincinnati, Ohio 45221}
\affiliation{Department of Physics, Fu Jen Catholic University, Taipei}
\affiliation{The Graduate University for Advanced Studies, Hayama, Japan}
\affiliation{University of Hawaii, Honolulu, Hawaii 96822}
\affiliation{High Energy Accelerator Research Organization (KEK), Tsukuba}
\affiliation{University of Illinois at Urbana-Champaign, Urbana, Illinois 61801}
\affiliation{Institute of High Energy Physics, Chinese Academy of Sciences, Beijing}
\affiliation{Institute of High Energy Physics, Vienna}
\affiliation{Institute of High Energy Physics, Protvino}
\affiliation{Institute for Theoretical and Experimental Physics, Moscow}
\affiliation{J. Stefan Institute, Ljubljana}
\affiliation{Kanagawa University, Yokohama}
\affiliation{Korea University, Seoul}
\affiliation{Kyungpook National University, Taegu}
\affiliation{Swiss Federal Institute of Technology of Lausanne, EPFL, Lausanne}
\affiliation{University of Maribor, Maribor}
\affiliation{University of Melbourne, Victoria}
\affiliation{Nagoya University, Nagoya}
\affiliation{Nara Women's University, Nara}
\affiliation{National Central University, Chung-li}
\affiliation{National United University, Miao Li}
\affiliation{Department of Physics, National Taiwan University, Taipei}
\affiliation{H. Niewodniczanski Institute of Nuclear Physics, Krakow}
\affiliation{Nippon Dental University, Niigata}
\affiliation{Niigata University, Niigata}
\affiliation{University of Nova Gorica, Nova Gorica}
\affiliation{Osaka City University, Osaka}
\affiliation{Osaka University, Osaka}
\affiliation{Panjab University, Chandigarh}
\affiliation{Peking University, Beijing}
\affiliation{RIKEN BNL Research Center, Upton, New York 11973}
\affiliation{Saga University, Saga}
\affiliation{University of Science and Technology of China, Hefei}
\affiliation{Seoul National University, Seoul}
\affiliation{Shinshu University, Nagano}
\affiliation{Sungkyunkwan University, Suwon}
\affiliation{University of Sydney, Sydney NSW}
\affiliation{Tata Institute of Fundamental Research, Bombay}
\affiliation{Toho University, Funabashi}
\affiliation{Tohoku Gakuin University, Tagajo}
\affiliation{Tohoku University, Sendai}
\affiliation{Department of Physics, University of Tokyo, Tokyo}
\affiliation{Tokyo Institute of Technology, Tokyo}
\affiliation{Tokyo Metropolitan University, Tokyo}
\affiliation{Tokyo University of Agriculture and Technology, Tokyo}
\affiliation{Virginia Polytechnic Institute and State University, Blacksburg, Virginia 24061}
\affiliation{Yonsei University, Seoul}
\author{M.-C.~Chang}\affiliation{Department of Physics, Fu Jen Catholic University, Taipei} 
  \author{K.~Abe}\affiliation{High Energy Accelerator Research Organization (KEK), Tsukuba} 
  \author{K.~Abe}\affiliation{Tohoku Gakuin University, Tagajo} 
  \author{I.~Adachi}\affiliation{High Energy Accelerator Research Organization (KEK), Tsukuba} 
  \author{H.~Aihara}\affiliation{Department of Physics, University of Tokyo, Tokyo} 
  \author{D.~Anipko}\affiliation{Budker Institute of Nuclear Physics, Novosibirsk} 
  \author{K.~Arinstein}\affiliation{Budker Institute of Nuclear Physics, Novosibirsk} 
  \author{T.~Aushev}\affiliation{Swiss Federal Institute of Technology of Lausanne, EPFL, Lausanne}\affiliation{Institute for Theoretical and Experimental Physics, Moscow} 
  \author{A.~M.~Bakich}\affiliation{University of Sydney, Sydney NSW} 
  \author{E.~Barberio}\affiliation{University of Melbourne, Victoria} 
% ...
\collaboration{The Belle Collaboration}

\noaffiliation

\begin{abstract}
We report the first observation of $B^0 \to J/\psi \ \eta$ decay.  
These results are obtained from a  data sample that contains 449
million $B\bar{B}$ pairs 
accumulated at the $\Upsilon(4S)$ resonance
with the Belle detector at the KEKB asymmetric-energy $e^+ e^-$ collider.
We observe a signal with a significance of 8.1$\sigma$
and obtain a branching fraction of (9.5 $\pm$ 1.7 (stat) $\pm$ 0.8 (syst))$\times 10^{-6}$.
\end{abstract}


\maketitle

\tighten

{\renewcommand{\thefootnote}{\fnsymbol{footnote}}}
\setcounter{footnote}{0}

\begin{figure}[htb]
\includegraphics[width=0.49\textwidth]{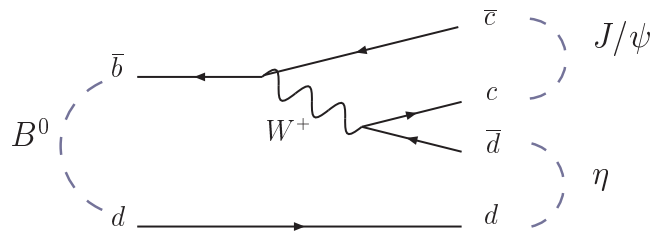}
\caption{Feynman diagram for the leading contribution to the decay $B^0 \to J/\psi \ \eta$.}
\label{fig:feyman}
\end{figure}

At the quark level,
the decay $B^0 \rightarrow J/\psi \eta$ proceeds primarily via a
$\bar{b} \rightarrow \bar{c} c \bar{d}$ transition.
As is apparent from its Feynman diagram (Fig.~\ref{fig:feyman}), this is a Cabibbo- and
color-suppressed decay, similar to $B^0 \rightarrow J/\psi \pi^0$
for which the branching fraction and $CP$ violation parameters have
been measured by the BaBar and Belle collaborations~\cite{CLEO-jpsipi0-brpaper,
BaBar-jpsipi0-brpaper, Belle-jpsipi0-brpaper, 
BaBar-jpsipi0-cppaper,Belle-jpsipi0-cppaper,
BaBar-jpsipi0-up}.
The best existing limit for the branching fraction for $B^0 \rightarrow J/\psi \eta$  comes from BaBar~\cite{BaBar-best-limit-jpsieta}
and is based on 56 million $B\bar{B}$ pairs.

If factorization and flavor-SU(3) symmetric coefficients for the
$\bar{d}d$ content are assumed,  the branching fraction for $B^0 \rightarrow J/\psi \eta$
decay is expected to be comparable to that for $B^0 \rightarrow J/\psi \pi^0$.
If the measured branching fraction is
significantly different from this expectation, it would imply the influence of large
final-state interactions or non-standard contributions.
In the latter case, precise measurements of $CP$ violation parameters
might also reveal unexpected phenomena.

In this letter, we report the first observation of the decay
$B^0 \rightarrow J/\psi \eta$~\cite{CC}.
The measurement
is based on a data sample that
contains 449 million $B\overline{B}$ pairs, 
collected  with the Belle detector at the KEKB asymmetric-energy
$e^+e^-$ (3.5 on 8~GeV) collider~\cite{KEKB}
operating at the $\Upsilon(4S)$ resonance.

The Belle detector is a large-solid-angle magnetic
spectrometer that
consists of a silicon vertex detector (SVD),
a 50-layer central drift chamber (CDC), an array of
aerogel threshold \v{C}erenkov counters (ACC), 
a barrel-like arrangement of time-of-flight
scintillation counters (TOF), and an electromagnetic calorimeter
comprised of CsI(Tl) crystals (ECL). All these detectors are located inside 
a superconducting solenoid coil that provides a 1.5~T
magnetic field.  An iron flux-return located outside of
the coil is instrumented to detect $K_L^0$ mesons and to identify
muons (KLM).  The detector
is described in detail elsewhere~\cite{Belle}.

The data used in this analysis was collected with 
two detector configurations. A 2.0 cm radius beampipe
and a 3-layer silicon vertex detector are used for the first sample
of 152 million $B\bar{B}$ pairs, while a 1.5 cm radius beampipe, a 4-layer
silicon detector and a small-cell inner drift chamber are used to record  
the remaining 297 million $B\bar{B}$ pairs~\cite{Ushiroda}.  
A GEANT-based Monte Carlo (MC) simulation is used to model the response of the
detector and determine efficiencies~\cite{GEANT}.

Charged tracks are selected based on the impact parameters relative to the
interaction point: $dr$ for the radial direction 
and $dz$ for the direction along the positron beam.
Our requirements are
$dr< 1$ cm and $|dz|< 5$ cm. 
From among the selected charged tracks, 
$e^+$ and $e^-$ candidates are identified 
by combining information from the ECL, the CDC ($dE/dx$)
and the ACC.
Identification of $\mu^+$ and $\mu^-$ candidates is based on track
penetration depth and the hit pattern in the KLM system. 
Charged pions are identified using the combined information
from the CDC ($dE/dx$), the TOF and the ACC.

Photon candidates are selected
from calorimeter showers not associated with charged tracks.
An energy deposition with a photon-like shape and with energy of at least 50 MeV in the barrel region or 100 MeV
in the endcap region is considered a photon candidate.
A pair of photons with an invariant mass
117.8 MeV/$c^2$ $< M_{\gamma\gamma}<$ 150.2 MeV/$c^2$
is considered as a $\pi^0$ candidate. This invariant mass region 
corresponds to a $\pm$3$\sigma$ interval around the $\pi^0$ mass,
where $\sigma$ is the mass resolution. 

We reconstruct $J/\psi$ mesons in the $\ell^+\ell^-$ decay channel 
($\ell$ = $e$ or $\mu$) and include up to one bremsstrahlung photon that is
within 50 mrad of each of the $e^+$ and $e^-$ tracks (denoted as $e^+e^-(\gamma)$). The invariant mass
 is required to be within $-$0.15 GeV/$c^2$ 
$<M_{ee(\gamma)}- m_{J/\psi}<$ 0.036 GeV/$c^2$ and $-$0.06 GeV/$c^2$ 
$<M_{\mu \mu}- m_{J/\psi}<$ 0.036 GeV/$c^2$, where $m_{J/\psi}$ denotes the 
nominal mass, $M_{ee(\gamma)}$ and $M_{\mu\mu}$ are the reconstructed 
invariant mass from $e^+e^-(\gamma)$ and $\mu^+\mu^-$, respectively.    
Asymmetric intervals are used to include part of the radiative tails.

Candidate $\eta$ mesons are reconstructed in the $\gamma\gamma$ and 
$\pi^+\pi^-\pi^0$ final states. We require the invariant mass to be
in the range 500 MeV/$c^2$ $< M_{\gamma\gamma} <$ 575 MeV/$c^2$ ($\pm$3$\sigma$) and
535 MeV/$c^2$ $< M_{\pi^+\pi^-\pi^0} <$ 560 MeV/$c^2$ ($\pm$3$\sigma$).
In the $\gamma\gamma$ final state, 
we remove $\eta$ candidates if either of the daughter photons 
form a $\pi^0$ candidate when combined with any other photon in the event. 
Asymmetric decays are removed with the requirement 
$|E_{\gamma 1}-E_{\gamma2}|/(E_{\gamma 1}+E_{\gamma2})<0.8$,
where $E_{\gamma1}$ and $E_{\gamma2}$
are the energies of the two photons 
that form the $\eta$ candidate. 

We combine the $J/\psi$ and $\eta$ to form $B$ mesons. Signal candidates 
are identified by 
 two kinematic variables defined in the $\Upsilon(4S)$ center-of-mass (CM) 
frame: the beam-energy constrained mass, 
$M_{\rm bc}=\sqrt{E^2_{\rm beam}-p^{*2}_{B}}$, 
and the energy difference, $\Delta E= E^*_B-E_{\rm beam}$, 
where $p^*_B$ and $E^*_B$ are the momentum and energy  of the $B$ candidate 
and $E_{\rm beam}$ is the run-dependent beam energy. 
To improve the $\Delta E$ resolution, 
the masses of the selected $\pi^0$, $\eta$ and $J/\psi$ candidates are constrained to 
their nominal masses using mass-constrained kinematic fits.
In addition, vertex-constrained fits are applied 
to the $\eta \to \pi^+ \pi^- \pi^0$ and $J/\psi \to \ell^+ \ell^-$ candidates.
We retain events with $M_{\rm bc} >$ 5.2 GeV/$c^2$ and
$|\Delta E| <$ 0.2 GeV, and define a signal region to be
$5.27\, {\rm GeV}/c^2 < M_{\rm bc} < 5.29\, {\rm GeV}/c^2$,
$-0.10\, {\rm GeV} < \Delta E < 0.05\, {\rm GeV}$ ($\eta \to \gamma \gamma$)
or $|\Delta E| < 0.05\, {\rm GeV}$ ($\eta \to \pi^+\pi^-\pi^0$).
In events with more than one $B$ candidate, 
usually due to multiple $\eta$ candidates,
the $B$ candidate with the minimum $\chi^2$ value from
the mass- and vertex-constrained fit is chosen.

To suppress the
combinatorial background dominated by the two-jet-like 
$e^+e^- \to q \bar{q}$ ($q=u,d,s$) continuum process, we remove events
with the ratio of second to zeroth 
Fox-Wolfram moments $R_2 >$ 0.4~\cite{SFW}. 
This requirement is determined using a figure-of-merit
$N_S/\sqrt{N_S+N_B}$, where $N_{S}$  is the number of expected signal
events and $N_B$ is the number of background events.
For $N_S$, we use a MC simulation with the  assumption that $\mathcal{B}(B^0 \to J/\psi \ \eta)$ is
$6 \times 10^{-6}$~\cite{br-assume}.  For $N_B$, we
use a sample of continuum background MC that corresponds to
an integrated luminosity that is about three times that of the data sample
and normalized to the number of events in an
$M_{\rm bc}-\Delta E$ sideband (5.2 GeV/$c^2$ $<M_{\rm bc}<$ 5.26 GeV/$c^2$ and
0.1 GeV $< \Delta E <$ 0.2 GeV).  
This requirement on $R_2$ eliminates 88\% of the continuum
background and retains 97\% of the signal.

After the above continuum suppression, the background is dominated by
 $B\bar B$ events with $B$ decays to $J/\psi$.
We use a MC sample corresponding to $3.86 \times 10^{10}$
generic $B\bar{B}$ decays that includes all known
 $B^0 \to J/\psi X$ processes  
to investigate these backgrounds. 
We find that the dominant backgrounds come from 
$B^0\to J/\psi K_L$ (25.1\%),
$B^{\pm} \to J/\psi K^{*\pm}(892)$ (19.3\%), 
$B^0 \to J/\psi K^{*0}(892)$ (14.5\%),
$B^0 \to J/\psi K_S$ (14.1\%),
$B^0 \to J/\psi \pi^0$ (8.4\%), 
and a few other exclusive $B \to J/\psi \ X$ decay modes.
These backgrounds peak in the $M_{\rm bc}$ distributions
but not in $\Delta E$.  Thus,
combinatorial backgrounds  from  $B \to J/\psi \ X$ decays
do not affect the signal yield extracted
from the $\Delta E$ distribution.

Signal yields and background levels are determined by fitting the distributions
in $\Delta E$ for candidates in the $M_{\rm bc}$ signal region.  
The $\Delta E$ distribution is fitted with a function
that is the sum of a Crystal Ball line function~\cite{CBLine} and 
two Gaussians for signal, and a second-order 
polynomial 
 for background.
The probability density functions (PDF) that describe the signal and 
background shapes are determined from MC simulations. 
We use the MC sample for this process to
 determine the PDF for the total background. 

We use the decay $B^+ \to J/\psi K^{*+}(K^{*+} \to K^+ \pi^0)$ 
as a control sample
to correct for differences  between data and MC for 
the fitted mean and width values of the $\Delta E$ signal peak.
For this selection, we require the helicity angle in the $K^{*+} \to K^+ \pi^0$
decay, {\it i.e.} the angle between the $\pi^0$ and the negative of the $B^+$ momenta
in the $K^*$ rest frame, to be less than $86$ degrees.  
This requirement primarily selects events with a high momentum $\pi^0$
 and produces a control-sample $\Delta E$ distribution that is similar
to that for our signal decay mode. 
The signal PDF is modified based on the results for
the control sample:
the mean value of $\Delta E$ is shifted by $-$3.81 $\pm$ 0.65 MeV
and the width is scaled by a factor of 0.99 $\pm$ 0.04. 

\begin{figure}[htb]
\includegraphics[width=0.45\textwidth]{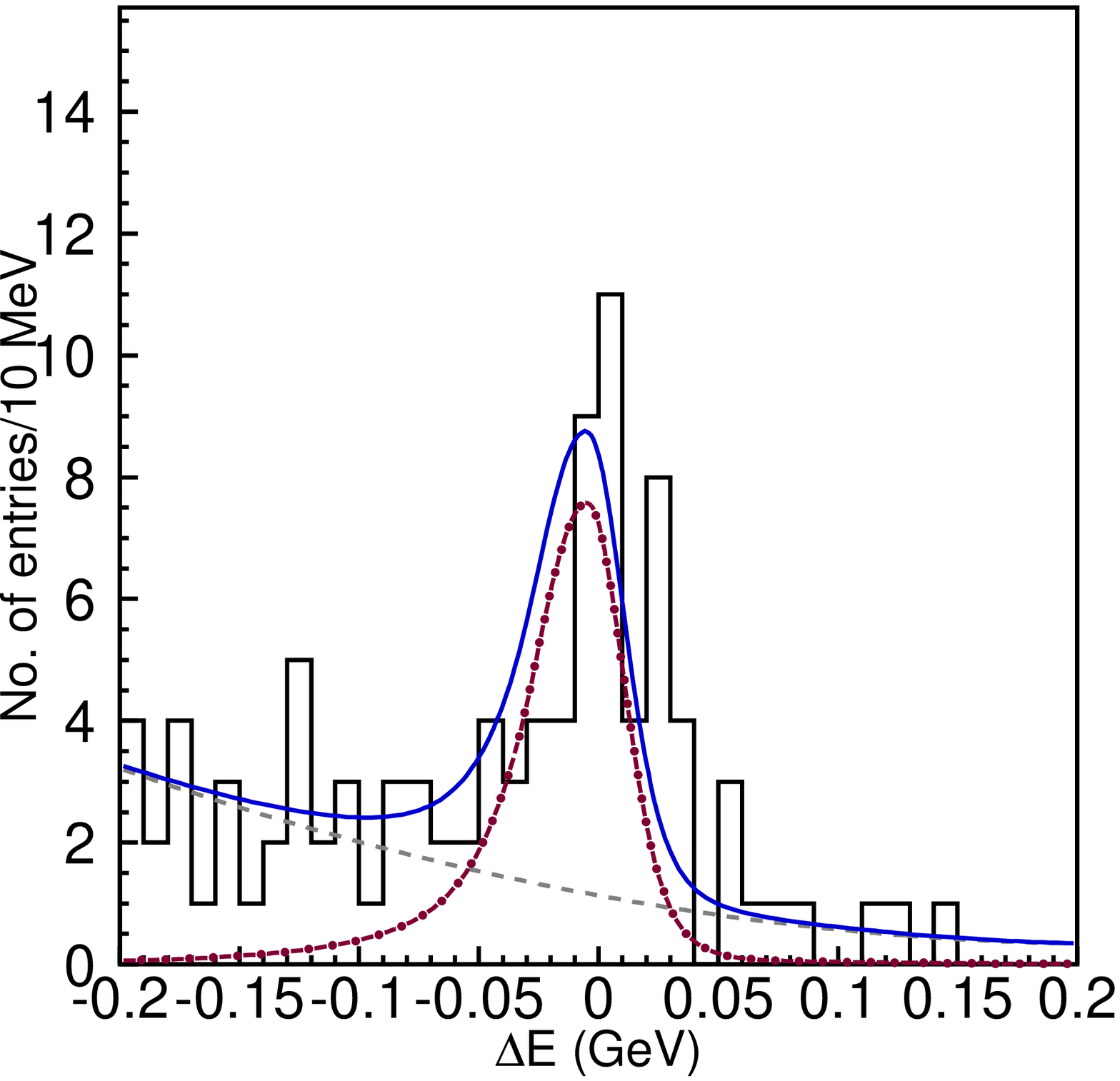}
\includegraphics[width=0.45\textwidth]{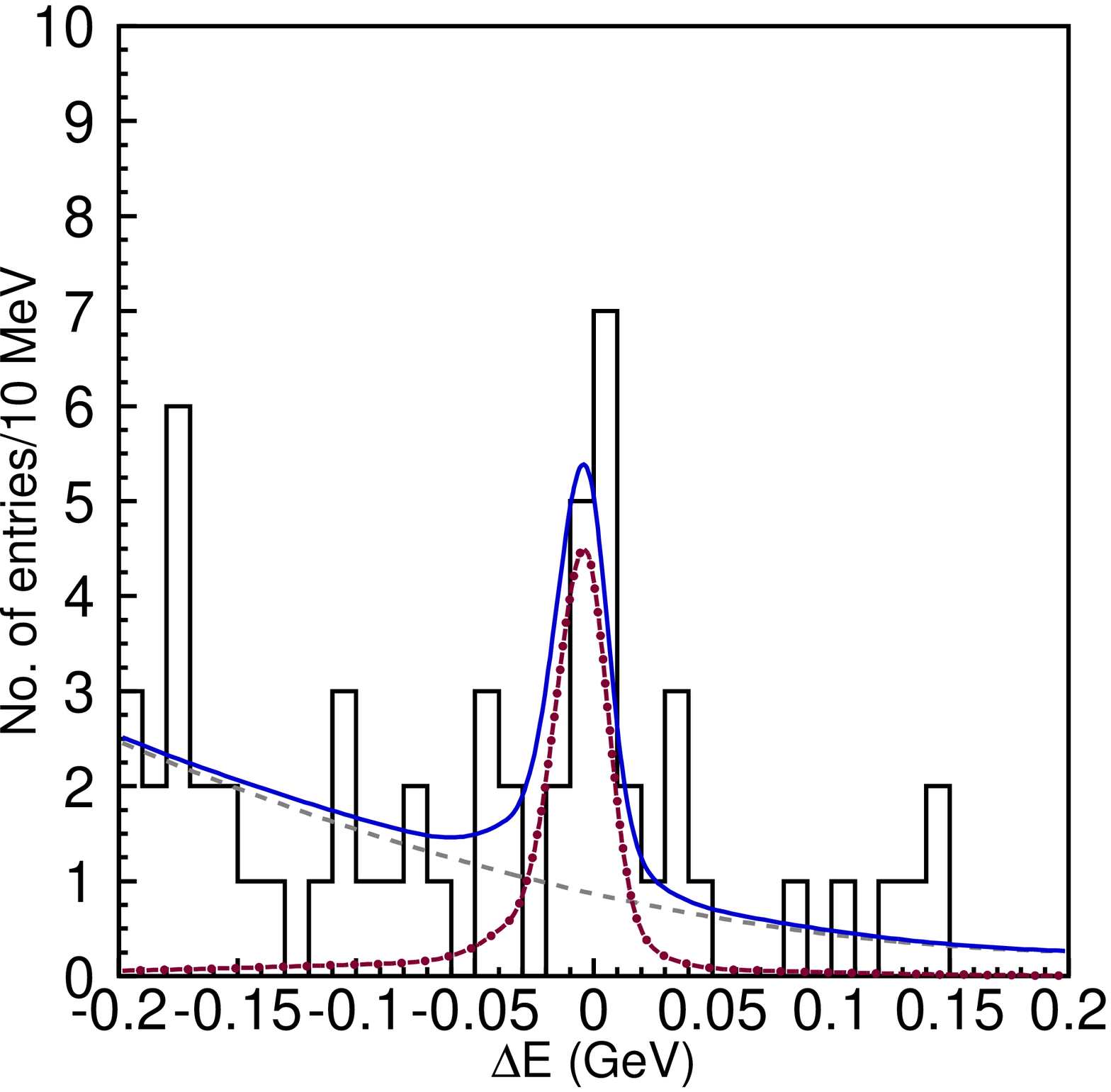}
\caption{$\Delta E$ distributions for the decay $B^0 \to \ J/\psi \eta$. 
The plot on the top shows $B^0 \to \ J/\psi \eta(\eta\to\gamma\gamma)$ while
 $B^0 \to \ J/\psi \eta(\eta\to\pi^+\pi^-\pi^0)$ is on the bottom.
The curves show the signal (dash-dot lines) and background (dashed
 lines) 
contributions 
as well as the overall fit (solid lines).}
\label{fig:defit}
\end{figure}

There are 98 and 58 events in the $M_{\rm bc}$ signal region for the 
$\eta \to \gamma \gamma$ 
and $\eta \to \pi^+ \pi^- \pi^0$ modes, respectively.
We determine the signal content by performing an unbinned 
extended maximum-likelihood fit to the candidate data events.
The unbinned extended maximum-likelihood function is 
\begin{equation}
\mathcal{L} = \frac{e^{-(N_S+N_B)}}{N!} \prod_i^N [N_S P_S(\Delta E_i)+N_B P_B(\Delta E_i)] \, ,
\end{equation}
where $N$ is the total number of candidate events, $N_S$ and $N_B$ denote the number of signal and $B\bar{B}$ background events,
$P_S(\Delta E_i)$ and $P_B(\Delta E_i)$ denote the signal and background $\Delta E$ PDFs respectively, and $i$ is the event index. 
Separate fits to the $B^0 \to J/\psi \eta (\gamma\gamma)$ and 
$B^0 \to J/\psi \eta (\pi^+\pi^-\pi^0)$ candidate samples give signal
 yields of  $43.1\pm8.9$  
and $16.6\pm5.8$ events, respectively. 
The results of the fits to the data are shown in Fig.~\ref{fig:defit}.
These signal yields correspond to branching fractions of  $(9.6\pm1.9)\times 10^{-6}$  
and $(10.0\pm3.5)\times 10^{-6}$, respectively.
We fit the two samples simultaneously, constraining the results to a common
branching fraction, and obtain 59.7 $\pm$ 10.5 events.
The combined branching fraction $(9.5\pm1.7)\times 10^{-6}$. 

We correct for discrepancies of the signal detection efficiency between
 data and 
MC using control samples.  The $J/\psi \to \mu^+\mu^-$ and 
$D^{*+} \to D^0 \pi^+$ $(D^0 \to K^-\pi^+)$ events are used
as control samples to correct for the muon identification and pion 
identification, respectively.
We find that the MC overestimates the efficiencies for
$J/\psi \to \mu^+\mu^-$ and $\eta \to \pi^+ \pi^- \pi^0$ by 8.5\% and 1.8\%,
respectively.

A 4.8\% systematic error associated with uncertainties in the signal
and background PDFs is estimated by comparing the fit results for the cases when the polynomial
parameters are fixed by either MC or data in the $M_{\rm bc}$ sideband region, 
and from changes that result from varying
each parameter by one standard deviation.
For the contribution from the corrections to the lepton identification
 efficiencies, we use a
control sample of $J/\psi \to \ell^+\ell^-$, which indicates uncertainties of 2.7\% per 
electron track and 1.2\% per muon track.  By weighting the relative
 contributions from $J/\psi \to e^+e^-$ and $J/\psi \to \mu^+\mu^-$,
we assign a 3.9\% systematic error. 
The systematic error from the pion identification correction
is 0.7\% per track; this is determined from a study of the
$D^{*+}\to D^0 \pi^+$ ($D^0 \to K^- \pi^+$) control sample.  
Since this applies only to the $\eta \to \pi^+ \pi^- \pi^0$ decay mode, 
the effective systematic error is 0.4\%. 
The systematic error due to the track finding efficiencies for the Belle
 detector is 1.0\% per charged lepton and 1.3\% per charged pion.  
This contributes 2\% per $J/\psi$ and 0.7\% per $\eta$, and the total of 2.7\%
(from linearly adding) is included.
A 4.1\% systematic error due to $\pi^0$ detection is determined
from a comparison of the  data and MC ratios for a large sample of
$\eta \to \pi^+ \pi^- \pi^0$ and $\eta \to 3 \pi^0$ decays. 
Since $\eta \to \gamma \gamma$ is similar to $\pi^0$ decay,
we also assign 4.1\% as the systematic error for $\eta \to \gamma \gamma$
reconstruction.
The errors on the branching fractions for 
$J/\psi \to \ell^+ \ell^-$,  $\eta \to \gamma \gamma$, 
and $\eta \to \pi^+ \pi^- \pi^0$ are 1.0\%, 0.7\%, and 1.8\%,
 respectively~\cite{PDG}. These errors contribute 1\% and 1.1\%
 systematic errors due to ${\cal B}$($J/\psi \to \ell^+ \ell^-$) and 
${\cal B}$($\eta \to \gamma \gamma$ and $\pi^+\pi^-\pi^0$), respectively. 
We assign a systematic error of 1.2\% due to the uncertainty in the
 number of $B \bar B$ pairs.

The systematic errors are summarized in Table~\ref{tab:sys}. 
The total uncertainty is the quadratic 
   sum of each term. The detection efficiencies 
   and branching fractions are listed in Table~\ref{tab:br}.
The statistical significance of the observed signal,
defined as $\sqrt{-2\ln(\mathcal{L}_0/\mathcal{L}_{\rm max})}$
where $\mathcal{L}_{\rm max}$ ($\mathcal{L}_0$) 
denotes the likelihood value at the maximum 
(with the signal yield fixed at zero), is 8.1$\sigma$.

\begin{table}[htb]
\caption{Systematic errors for the combined branching fraction.}
\label{tab:sys}
\begin{tabular}
{lcc}
\hline \hline
  & \multicolumn{2}{c}{Systematic errors(\%)}  \\
\hline
PDFs &  \multicolumn{2}{c}{4.8} \\
Lepton identification &\multicolumn{2}{c}{3.9} \\
Charged pion identification & \multicolumn{2}{c}{0.4}\\
Tracking (lepton and charged pion)&  \multicolumn{2}{c}{2.7}  \\
$\eta \to \gamma \gamma$, $\pi^0$ selection & \multicolumn{2}{c}{4.1} \\
${\cal B}$($J/\psi \to \ell^+ \ell^-$) & \multicolumn{2}{c}{1.0} \\
${\cal B}$($\eta \to \gamma \gamma$ and $\pi^+ \pi^- \pi^0$)&\multicolumn{2}{c}{1.1} \\
$N_{B\bar{B}}$ & \multicolumn{2}{c}{1.2} \\
\hline
Total & \multicolumn{2}{c}{8.1} \\
\hline \hline
\end{tabular}
\end{table}
\begin{table}[htb]
\caption{Detection efficiencies and branching fractions.}
\label{tab:br}
\begin{tabular}
{lcc}
\hline \hline
Mode & Eff.(\%) & BF($10^{-6}$) \\
\hline
$ B^0 \to J/\psi \eta(\gamma\gamma)$& 21.9 & 9.6 $\pm$ 1.9\\
$B^0 \to J/\psi \eta(\pi^+\pi^-\pi^0)1$ & 13.8 &10.0 $\pm$ 3.5 \\
$ B^0 \to J/\psi \eta$ (combined)& & 9.5 $\pm$ 1.7 $\pm$ 0.8 \\
\hline \hline
\end{tabular}
\end{table}

In summary, 
the first observation of the decay 
   $B^0 \to J/\psi \eta$ is reported. We observe 59.7 $\pm$ 10.5 
   signal events with 8.1$\sigma$ significance in a 449 million $B\bar{B}$ pair data sample at the $\Upsilon(4S)$ resonance.
   The measured branching fraction is 
$\mathcal{B} (B^0 \to J/\psi \ \eta)$ =
$(9.5 \pm 1.7 \pm 0.8)\times 10^{-6}$,
where the first error is statistical and the second is systematic.  
The measured branching fraction is consistent with the theoretical expectation~\cite{theory-1993} and
comparable to the world-average value for the $J/\psi \pi^0$ mode: $(2.2\pm 0.4)\times 10^{-5}$~\cite{PDG}. There is no indication of either large final state interactions or
non-standard model contributions.

We thank the KEKB group for excellent operation of the
accelerator, the KEK cryogenics group for efficient solenoid
operations, and the KEK computer group and
the NII for valuable computing and Super-SINET network
support.  We acknowledge support from MEXT and JSPS (Japan);
ARC and DEST (Australia); NSFC and KIP of CAS (contract
No.~10575109 and IHEP-U-503, China); DST (India); the BK21
program of MOEHRD, and the
CHEP SRC and BR (grant No. R01-2005-000-10089-0) programs of
KOSEF (Korea); KBN (contract No.~2P03B 01324, Poland); MIST
(Russia); ARRS (Slovenia);  SNSF (Switzerland); NSC and MOE
(Taiwan); and DOE (USA).


\begin{thebibliography}{99}

\bibitem{CLEO-jpsipi0-brpaper} 
P.~Avery {\it et al.} (CLEO Collab.),
Phys. Rev. D {\bf 62}, 051101 (2000).

\bibitem{BaBar-jpsipi0-brpaper}
B.~Aubert {\it et al.} (BaBar Collab.),
Phys. Rev. D {\bf 65}, 032001 (2002).

\bibitem{Belle-jpsipi0-brpaper} 
K.~Abe {\it et al.} (Belle Collab.),
Phys. Rev. D {\bf 67}, 032003 (2003).

\bibitem{BaBar-jpsipi0-cppaper}
B.~Aubert {\it et al.} (BaBar Collab.),
Phys. Rev. Lett. {\bf 91}, 061802 (2003).

\bibitem{Belle-jpsipi0-cppaper} 
S.~U.~Kataoka {\it et al.} (Belle Collab.),
Phys. Rev. Lett. {\bf 93}, 261801 (2004).

\bibitem{BaBar-jpsipi0-up}
B.~Aubert {\it et al.} (BaBar Collab.),
Phys. Rev. D {\bf 74}, 011101 (2006).

\bibitem{BaBar-best-limit-jpsieta}
B.~Aubert {\it et al.} (BaBar Collab.), 
Phys. Rev. Lett. {\bf 91}, 071801 (2003).

\bibitem{CC} Inclusion of charge conjugate modes is implied.
\bibitem{KEKB}
S.~Kurokawa and E.~Kikutani, Nucl. Instr. and. Meth. A {\bf 499}, 1 (2003),
and other papers included in this volume.

\bibitem{Belle}
A.~Abashian {\it et al.} (Belle Collab.),
Nucl. Instr. and Meth. A {\bf 479}, 117 (2002).

\bibitem{Ushiroda} Z. Natkaniec (Belle SVD2 Group),
Nucl. Instr. and Meth.A {\bf 560}, 1 (2006). 

\bibitem{GEANT} R. Brun {\it et al.}, GEANT 3.21, CERN DD/EE/84-1,1984.

\bibitem{SFW}
 The Fox-Wolfram moments were introduced in
 G.~C.~Fox and S.~Wolfram, Phys. Rev. Lett. {\bf 41}, 1581 (1978).

\bibitem{br-assume} For this optimization, we assume a branching fraction for
          $J/\psi \ \eta$ that is about one third that for $J/\psi \ \pi^0$. 

\bibitem{CBLine} DESY Internal Report, DESY F31-86-02(1990).

\bibitem{PDG} W.-M.~Yao {\it et al.}, (PDG), J. Phys. G {\bf 33}, 1 (2006).

\bibitem{theory-1993} A.~Deandrea {\it et al.}, Phys. Lett. B {\bf 318}, 549 (1993).

\end{thebibliography}
\end{document}